\documentclass[referee]{raa}
\usepackage{graphicx,times}
\usepackage{natbib}
\usepackage{amssymb,amsmath}
\usepackage{caption}
\bibpunct{(}{)}{;}{a}{}{,}

\usepackage[a4paper=true,dvipdfm=true,pagebackref=true]{hyperref}
\hypersetup{pdftitle = The title of my PDF, pdfauthor = My name, pdfsubject= The subject, pdfkeywords = keyword1 keyword2 keyword3}
\hypersetup{colorlinks = true, linkcolor = green, anchorcolor = red, citecolor = blue, filecolor = red, pagecolor = red, urlcolor = red}

\begin{document}

   \title{Constraints on the ultracompact minihalos using neutrino signals from the gravitino dark matter decay
$^*$
\footnotetext{\small $*$ Supported by the National Natural Science Foundation of China.}
}

 \volnopage{ {\bf 2012} Vol.\ {\bf X} No. {\bf XX}, 000--000}
   \setcounter{page}{1}

   \author{Yunlong Zheng\inst{1}, Yupeng Yang\inst{1,2,4,5}, Mingzhe Li\inst{3}, Hongshi Zong\inst{1,5}
   }

   \institute{ Department of Physics, Nanjing University, Nanjing 210093,China; {\it yyp@chenwang.nju.edu.cn}\\
        \and
       Department of Astronomy, Nanjing University, Nanjing 210093, China \\
        \and
     Interdisciplinary Center for Theoretical Study, University of Science and Technology of China, Hefei 230026, China \\
       \and
     Collage of Physics and Electrical Engineering, Anyang Normal University, Anyang 455000, China \\
       \and 
       Joint Center for Particle, Nuclear Physics and Cosmology, 
       Nanjing University-Purple mountain Observatory, Nanjing 210093, China\\
\vs \no
}

\abstract{ Ultracompact dark matter minihalos (UCMHs) would be formed during the earlier 
universe if there were large density perturbations. If the dark matter can decay
into the standard model particles, such as neutrinos, these objects would become the potential 
astrophysical sources and could be detected 
by the related instruments, such as IceCube. 
In this paper, we investigate the neutrino signals 
from the nearby UCMHs due to the gravitino dark matter decay and compare these signals 
with the background neutrino flux which is mainly from the atmosphere to get 
the constraints on the abundance of UCMHs. 
}

   \authorrunning{Y.-L. Zheng  Y.-P. Yang et al. }
   \titlerunning{Constraints on the abundance of ultracompact minihalos 
using neutrino signals}
   \maketitle

%
\section{Introduction}           
\label{sect:intro}

Structure formation is one of important research fields in cosmology. 
According to the theory, the present cosmic 
structures originated from the earlier density perturbations with the amplitude 
$\delta\rho/\rho\sim10^{-5}$ and this has been confirmed by many observations. 
On the other hand, primordial black holes would be formed if there were large 
density perturbations ($\delta\rho/\rho>0.3$) in the earlier universe (\citealt{PBH}). Recently, 
\cite{Ricott+Gould} found that if the amplitude of density perturbations was 
between above values a new kind of dark matter structures named 
ultracompact dark matter minihalos (UCMHs) would be formed. 
Compared with the classical dark matter halos, the formation time of these objects 
is earlier ($z\sim 1000$) and the density profile is steeper ($\rho(r) \sim r^{-2.25}$). 
If dark matter is in the form of 
weakly interacting massive particles (WIMPs), such as the neutralino, they can annihilate 
into the standard model particles, such as photons, positrons or neutrinos 
(\citealt{dm_1,dm_2}). Moreover, because 
the dark matter annihilation rate is proportional to the square of number density, 
the UCMHs would become one kind of the potential astrophysical sources 
(\citealt{scott_prl,scott_prd,prd_positron,prd_neutrino}). 
Besides the annihilation, decay is another important approach to detect dark matter signals. 
This is especially crucial for those dark matter candidates which do not annihilate. 
A famous example is the gravitino dark matter which in some supergravity models 
is the lightest supersymmetric particle 
(\citealt{dm_2}). Although compared with the annihilation 
the decay rate is proportional to the number density 
instead of the number density square, the decay is still very important for 
the cosmological probes for the dark matter particles which do not have the annihilation channels. 
\cite{epl} have investigated the gamma-ray flux 
from nearby UCMHs due to the dark matter decay. Through comparing with the observations 
they obtained the constraints on the abundance of UCMHs for 
different decay channels, lifetimes and density profiles of Milky Way. 
They found that the strongest constraint comes from the $b\overline{b}$ channel with 
the dark matter mass $m_{\chi}=100\mathrm{GeV}$, 
the fraction of UCMHs is $f_{\mathrm{UCMHs}}\sim5\times10^{-5}$.

Besides the high energy photons the other kind of important products of 
dark matter decay are neutrinos and they usually accompany with photons. 
The advantage of neutrino detection is that neutrinos can propagate in the space 
without attenuation due to its very weakly interaction with other particles. 
Therefore, comparing with other particles (e.g. electrons and positrons) 
the orientation of corresponding sources can be confirmed directly. 
When neutrinos propagate through 
the medium, such as the ice, muons ($\mu$) can be produced by 
the charged current interaction and detected by the Cherenkov radiation detector. 
Because the neutrino signals accompanying with the production of gamma-ray, 
the study on neutrino signals would be complementary to the gamma-ray observations 
especially for the larger dark matter mass and lepton channels 
(\citealt{com_neu_1,com_neu_2,com_neu_3}). 
In this paper, we will investigate the neutrino signals from UCMHs due to the 
dark matter decay. As we have not observed any excess of neutrino flux 
comparing with the atmospheric neutrino flux, we get the 
constraints on the abundance of UCMHs.

This paper is organised as follows. The neutrino flux from nearby UCMHs due to 
dark matter decay are studied in section 2. In section 3, we obtain 
the constraints on the fraction of UCMHs. We conclude with discussions in section 4.

\section{The basic quality of UCMHs and potential neutrino signals due to dark matter decay}
\label{sect:Obs}

After the formation of UCMHs, the dark matter particles will
be accreted by the radial infall. The density profile of UCMHs can be obtained through the simulation (\citealt{Ricott+Gould,scott_prl}), 

\begin{equation}
    \rho(r,z)=\frac{3f_{\chi}M_\mathrm{UCMHs}(z)}{16\pi R(z)^{\frac{3}{4}}r^{\frac{9}{4}}}
\end{equation}
where $M_{\mathrm{UCMHs}}\mathrm{(z)} = M_{i}\frac{1+z_\mathrm{eq}}{1+z}$ is the mass of UCMHs 
and $M_i$ is the mass within the scale of perturbations, 
$R(z)=0.019(\frac{1000}{z+1})(\frac{M_\mathrm{UCMHs}(z)}{M_{\odot}})^{\frac{1}{3}}\mathrm{pc}$ 
is the radius of UCMHs at redshift $z$ and $f_{\chi}=\frac{\Omega_\mathrm{CDM}}
{\Omega_\mathrm{CDM}+\Omega_\mathrm{b}}=0.83$ (\citealt{wmap_7}).

The muons produced in the detector through the charged current interactions 
are called contained events. The flux can be written as (\citealt{erkoca,prd_neutrino})

\begin{equation}
    \frac{d\phi_{\mu}}{dE_{\mu}}=\int^{m_{\chi}}_{E_{\mu}}dE_{\nu}\frac{d\phi_{\nu}}{dE_{\nu}}
    \frac{N_{A} \rho}{2}\left(\frac{d\sigma^{P}_{\nu}(E_{\nu},E_{\mu})}{dE_{\mu}}+(p\rightarrow n)\right)+(\nu\rightarrow\overline{\nu}),
\end{equation}
where $N_{A}=6.022\times10^{23}$ is Avogadro's number and $\rho$ is the density of medium. 
$d\sigma^{p,n}_{\nu,\overline{\nu}}/dE_{\mu}$ are the weak scattering charged-current 
cross sections for neutrino and antineutrino scattering with protons and 
neutrons. $d\phi_{\nu}/dE_{\nu}$ is the differential flux of neutrino from UCMHs due to dark matter decay, 

\begin{equation}
    \frac{d\phi_{\nu}}{dE_{\nu}}= \frac{1}{d^{2}} \times \frac{1}{m_{\chi}\tau}\left(\sum B_{F}\frac{dN^{F}_{\nu}}{dE_{\nu}}\right)\int\rho(r)r^{2}dr,
\end{equation}
where $dN_{\nu}/dE_{\nu}$ is the neutrino number per dark matter decay, $B_{F}$ is the branching ratio 
of every decay channel. $m_{\chi}$ and $\tau$ are the dark matter mass and lifetime respectively. 
$d$ is the distance of UCMHs. In this paper, we consider gravitino ($\psi_\frac{3}{2}$) as the dark matter 
decay model. The gravitino is the lightest supersymmetric particle and 
they can decay into standard model particles in the presence of R-parity breaking (\citealt{gravitino}). 
There are three-body and two-body decay models for the gravitino particles and 
the three-body decay models can supply one of the explainations of positron excess observed by the PAMELA 
and Fermi experiments (\citealt{gravitino_for_pamela}). In this paper, we mainly 
consider this model, $\psi_\frac{3}{2} \to l^{+}l^{-}\nu$. 
Here, $l$ could be $\mu$ or $\tau$. Because the observation and identification 
of the $\nu_\mu$ is easier than that of the $\nu_\tau$, so we mainly consider 
the $\mu$($\nu_\mu$) channel. On the other hand, the second muon neutrinos 
can also be produced through the decay of $\mu$ which 
are from the three-body decay. Although the contribution of these process 
are much smaller than the primary case, we also include them in our calculations. For the neutrino 
spectrum of these decay channels, we use the forms which have been given in the Refs. (\cite{Erkoca:2010vk}). 


Besides the contained events, the muons flux produced before arriving at the detector 
is called upward events (\citealt{erkoca,yq_1}),

\begin{equation}
    \frac{d\phi_{\mu}}{dE_{\mu}}=\int^{m_{\chi}}_{E_{\mu}}dE_{\nu} \frac{d\phi_{\nu}}{dE_{\nu}}
    \frac{N_{A} \rho}{2}\left(\frac{d\sigma^{P}_{\nu}(E_{\nu},E_{\mu})}{dE_{\mu}}+(p\rightarrow n)\right)R(E_{\mu})+(\nu\rightarrow\overline{\nu}),
\end{equation}
where $R(E_{\mu})$ is the muon range or stopping distance at which muons 
can propagate in matter until their energy is below the threshold of the detector 
$E_{\mu}^{th}$ and it is given by $R(E_{\mu})=\frac{1}{\beta\rho}
\ln(\frac{\alpha+\beta E_{\mu}}{\alpha+\beta E_{\mu}^{th}})$ (\citealt{neutrino_r_mu}), 
with $\alpha$ corresponding to the ionization energy loss and $\beta$ 
accounts for the bremsstrahlung pair production and photonuclear interactions. 

For the neutrino detection, the main background is from the atmosphere. 
The angle-averaged atmospheric flux (ATM) is in the form of 
(in units of $\mathrm{Gev^{-1}km^{-2}yr^{-1}sr^{-1}}$) (\citealt{atm})

\begin{equation}
   \left(\frac{d\phi_{\nu}}{dE_{\nu}d\Omega}\right)_\mathrm{ATM}=
N_{0}E_{\nu}^{-\gamma-1}\times\left(\frac{a}{bE_{\nu}}\ln(1+bE_{\nu})+\frac{c}{e E_{\nu}}\ln(1+eE_{\nu})\right)
\end{equation}
where $a = 0.018, b= 0.024, c = 0.0069, e = 0.00139$, $\gamma = 1.74$ and 
$N_0 = 1.95(1.35) \times 10^{17}$ for $\nu(\bar\nu)$.

\begin{figure}[h]
  \begin{minipage}[t]{0.495\linewidth}
  \centering
   \includegraphics[width=70mm,height=90mm]{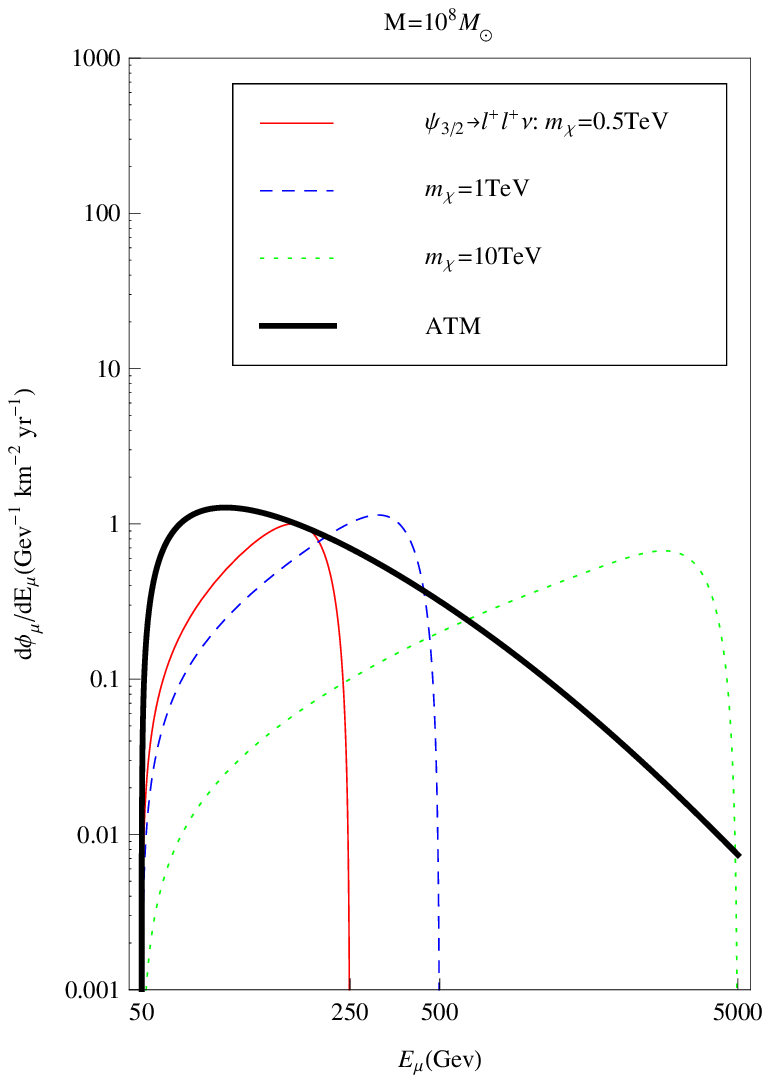}
  \end{minipage}%
  \begin{minipage}[t]{0.495\textwidth}
  \centering
   \includegraphics[width=70mm,height=90mm]{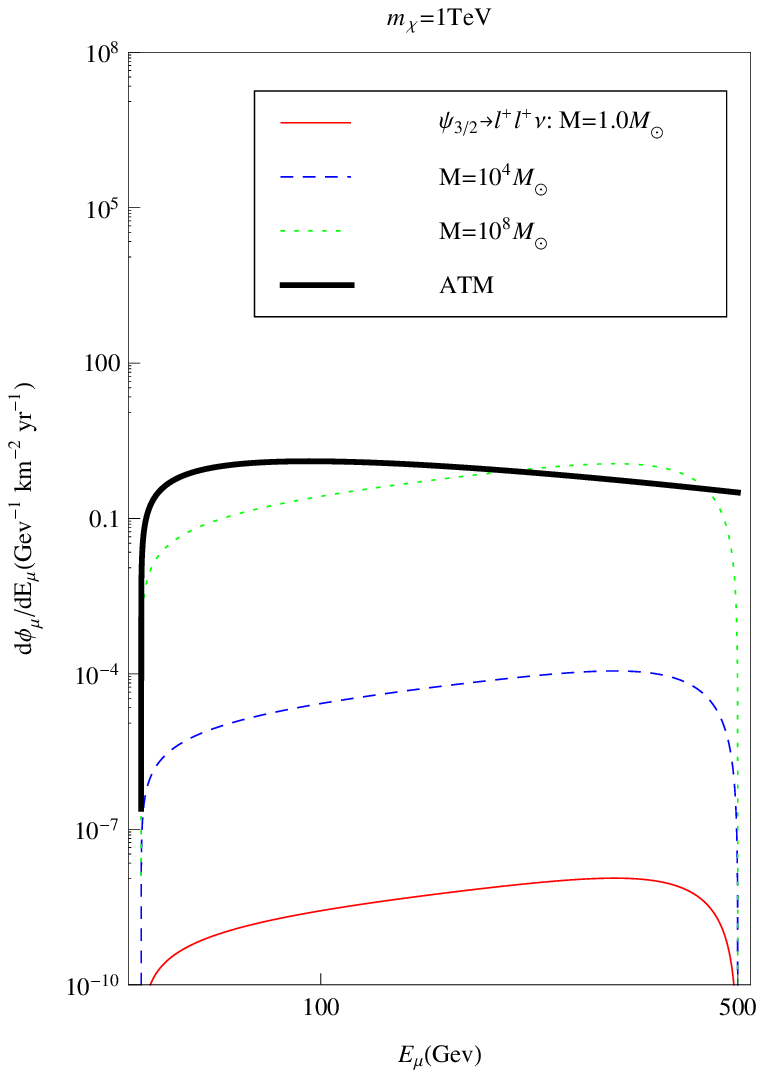}
  \end{minipage}%
\caption{\label{fig:up} The potential neutrino flux for the upward events from 
nearby UCMHs due to the gravitino decay. The distance of UCMHs is $d=1\mathrm{kpc}$ 
and the decay rate is $\Gamma=10^{-26}\mathrm{s^{-1}}$. Right: the neutrino 
flux for the fixed dark matter mass $m_{\chi} = 1\mathrm{TeV}$ and the mass 
of UCMHs $M_\mathrm{UCMHs} = 1\mathrm{M_\odot}$ (solid line), 
$10^{4}\mathrm{M_\odot}$ (long-dashed line), 
$10^{8}\mathrm{M_\odot}$ (dot-dashed line) from bottom to 
up respectively. 
Left: the neutrino flux for the fixed mass of UCMHs 
$M_\mathrm{UCMHs} = 10^{8}\mathrm{M_\odot}$ and the dark 
matter mass $m_{\chi} = 0.5\mathrm{TeV}$ (solid), $1\mathrm{TeV}$ (long-dashed line), 
$10\mathrm{TeV}$ (dot-dashed line) from left to right respectively. The neutrino flux 
from the atmosphere are also shown (thick solid line).}
\end{figure}

\begin{figure}[h]
  \begin{minipage}[t]{0.495\linewidth}
  \centering
   \includegraphics[width=70mm,height=90mm]{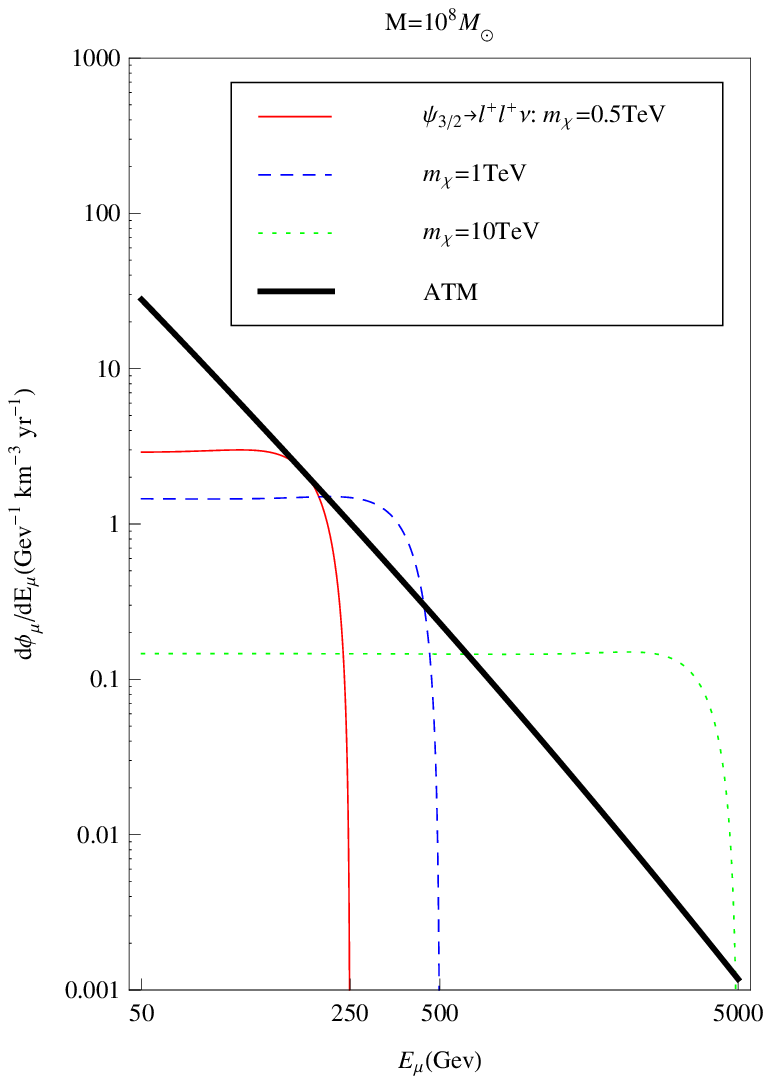}
  \end{minipage}%
  \begin{minipage}[t]{0.495\textwidth}
  \centering
   \includegraphics[width=70mm,height=90mm]{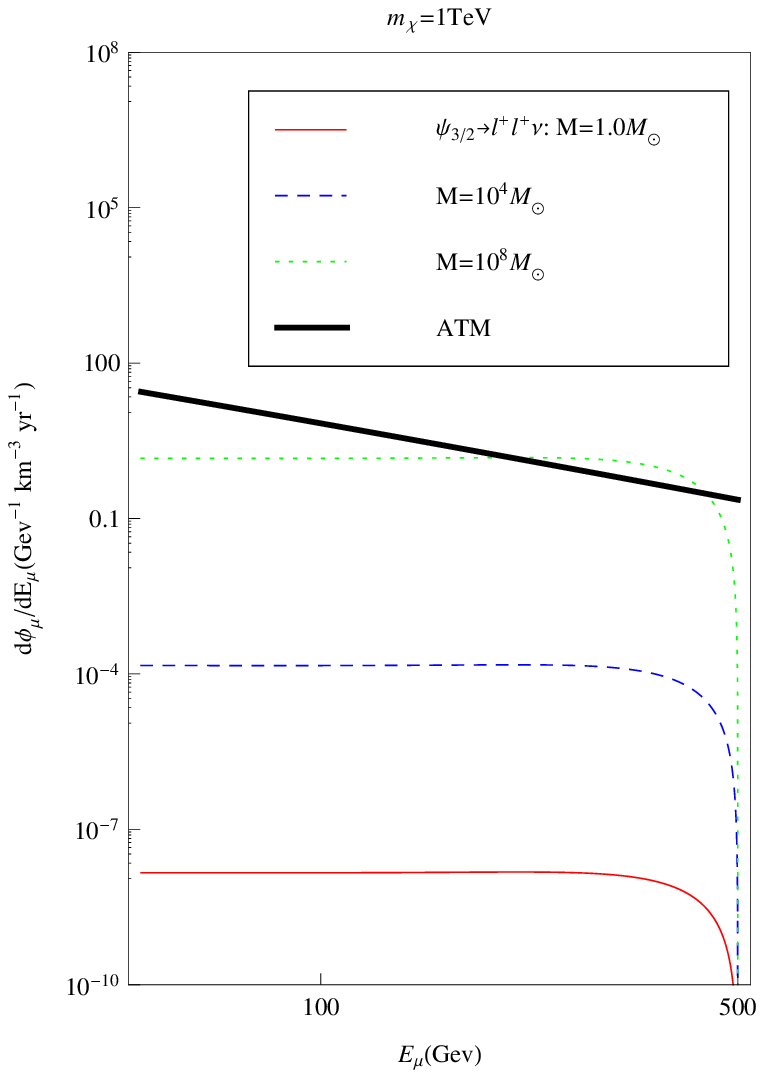}
  \end{minipage}%
\caption{\label{fig:con} The potential neutrino flux for the contained events from 
nearby UCMHs due to the gravitino decay. The parameters are the same as in the 
Fig.~\ref{fig:up}.}
\end{figure}

In the Figs.~\ref{fig:up} and~\ref{fig:con}, the neutrino flux from nearby 
UCMHs ($d_{\mathrm{UCMHs}} = 1 \mathrm{kpc}$) are shown for different dark matter mass. 
For these results, we set the dark matter decay rate as $\Gamma = 10^{-26} \mathrm{s^{-1}}$ 
and treat the distribution of neutrino flavors at earth as 
1:1:1 which is due to vacuum oscillation during the propagation. 
From these figures, it can be seen that for the fixed mass of UCMHs, the flux 
would excess the ATM for the larger dark matter mass. For the fixed mass 
of dark matter, the flux have excess the ATM for the bigger UCMHs. The similar results 
can also be found in our another work (\citealt{prd_neutrino}) where the neutrino 
signals from the UCMHs due to dark matter annihilation are considered.

\section{Constraints on the abundance of UCMHs}

After the formation of UCMHs, one of the important questions is about the 
abundance of them in our universe. According to above discussions, 
this issue can be solved through studying 
the present different observational results (\citealt{scott_prd,prd_neutrino,prd_positron,epl,Josan:2010vn,
Yang:2011jb,Yang:2011ef}). In the previous work, we get the constraints on the 
abundance of UCMHs considering the neutrino signals due to dark matter annihilation (\citealt{prd_neutrino}). 
In this section, we will investigate the fraction of UCMHs in the case of 
dark matter decay.

Following the previous works (e.g. \cite{scott_prd}), we assume that the UCMHs are distributed uniformly 
and the abundance is the same everywhere 
in our universe. The mass function is in the delta form which means that 
all of these objects have the same mass. Therefore, the fraction of UCMHs can be written 
as (\citealt{Josan:2010vn}) 

\begin{equation}
f_\mathrm{UCMH} = \frac{M_\mathrm{UCMH}}{M_\mathrm{DM,MW}(<d)}
\end{equation}
Where $M_\mathrm{DM,MW}(<d)$ is the dark matter mass of Milky Way within 
the distance $d$ and in this work we use the Navarro-Frenk-White profile 
for the dark matter halo.

\begin{figure}[h]
  \begin{minipage}[t]{0.49\linewidth}
  \centering
   \includegraphics[width=75mm,height=70mm]{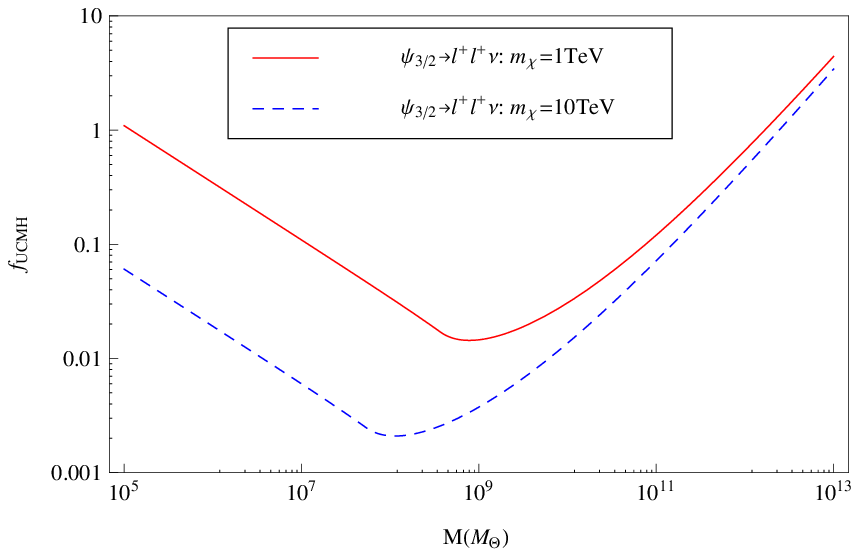}
  \end{minipage}%
  \begin{minipage}[t]{0.49\textwidth}
  \centering
   \includegraphics[width=75mm,height=70mm]{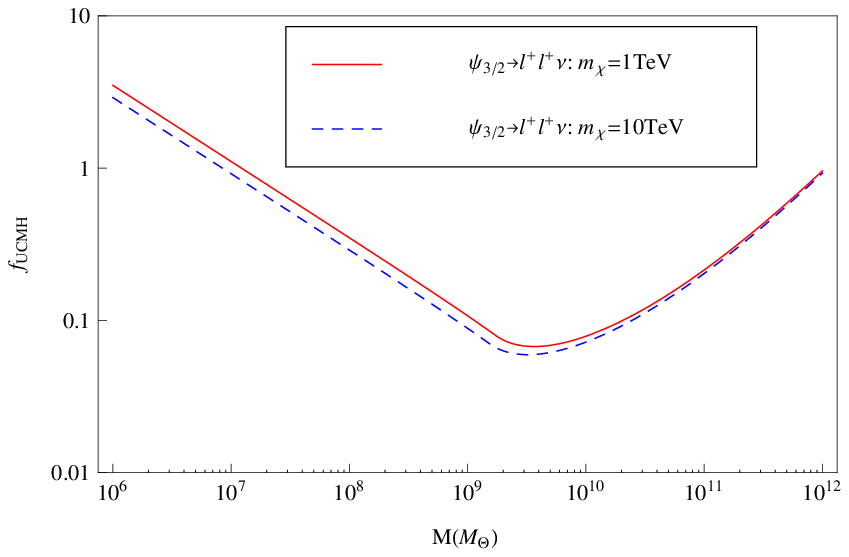}
  \end{minipage}%
\caption{\label{fig:cons} The constraints on the abundance of UCMHs 
using the potential neutrino signals, contained events(Right) and 
upward events(Left). We have chosen the dark matter mass $m_{\chi} = 
1\mathrm{TeV}$(solid line), $10\mathrm{TeV}$(dashed line) and 
set the decay rate as $\Gamma=10^{-26}\mathrm{s^{-1}}$.}
\end{figure}

In the section 2, it can be seen that the neutrino 
flux would excess the ATM for some cases. During the exposure times, such as 
ten years, considering the contamination of atmoshere background, 
the minimal number of muons from UCMHs for the fixed distance and 
$2\sigma$ statistic significance can be obtained as 
$N_\mathrm{UCMHs}/\sqrt{N_\mathrm{UCMHs}+N_\mathrm{ATM}} = 2$, 
where $N_\mathrm{{UCMHs}}$ can be obtained by integration
\begin{equation}
    N_\mathrm{UCMHs}=\int^{E_\mathrm{max}}_{E_{\mu}^{th}}\frac{d\phi_{\mu}}{dE_{\mu}}F_\mathrm{eff}(E_{\mu})dE_{\mu},
\end{equation}
where $F_\mathrm{eff}(E_{\mu})$ is the effective volume $V_\mathrm{eff}$ 
and effective area $A_\mathrm{eff}$ of the detector for 
the contained and upward events respectively. 
For the IceCube, we neglect the energy dependence and 
accept that $V_\mathrm{eff}=0.04 \mathrm{km^{3}}$, 
$A_\mathrm{eff}=1 \mathrm{km^{2}}$.  The final results are shown in the Fig.~\ref{fig:cons}.

From these results it can be seen that the strongest constraints is 
$f_\mathrm{UCMHs} \sim 2 \times 10^{-3}$ for the dark matter mass 
$m_\mathrm{DM} = 10 \mathrm{TeV}$ for the upward events. 
For the case of contained events, the constraints are weaker and the strongest 
constraints is $f_\mathrm{UCMHs} \sim 0.1$. So the upward events is much more 
competitive for the constraints on UCMHs and the limits will be stronger for 
the larger dark matter mass. On the other hand, comparing with the dark 
matter annihilation cases, these constraints are weaker. However, as we have mentioned 
in above sections that if the dark matter particles do not annihilate, dark matter 
decay will be important, so these results are still very significant.

\section{Discussions}

The UCMHs would be formed if there are large density perturbations during the 
earlier epoch and then its cosmological abundance become one of the important issuses. 
In the previous works, the main limits are from the research of the 
gamma-ray flux due to dark matter annihilation within UCMHs. In our recently 
works, we considered the constraints from the neutrino flux due to the 
dark matter annihilation and the gamma-ray flux due to the dark matter decay (\cite{epl,prd_neutrino}). 
In this work, we extended these works and investigated the neutrino flux from 
nearby UCMHs due to the gravitino particles decay. The decay styles of gravitino include 
two-body and three-body decays. The latter can also provide one of the explanations 
of the positrons (or positrons plus electrons) excess which have been observed recently 
by the PAMALE and Fermi. In this work, we mainly considered the three-body decay 
style. Most of this decay productions are leptons, therefore, the neutrinos 
will be plenty. We researched the neutrino flux from nearby UCMHs due to 
the dark matter decay and compared that with the signals from the atmosphere which 
is the main background of neutrino detection. We found that for the larger 
dark matter mass or the UCMHs the final flux would excess the ATM. These results 
are similar to the dark matter annihilation cases (\citealt{prd_neutrino}). 
On the other hand, because we have not observed any excess of neutrino flux 
from nearby unknown sources, so the abundance of UCMHs can be constrained. 
We considered ten years exposure times for neutrino observation and 
signals with 2$\sigma$ statistic significance to get the limits on the fraction 
of UCMHs. In this work, we also assumed that the distribution of these objects is uniform 
in the universe. We found that the strongest limits on the abundance of UCMHs is 
$f_\mathrm{UCMHs} \sim 10^{-3}$. One should note that these results depend 
on the dark matter mass. From the Fig.~\ref{fig:cons}, it can be seen that 
the limits will be stronger for the larger dark matter mass. Moreover, the final 
constraints are stronger for the upward events comparing with the contained events. 
The other important parameters is the dark matter decay rate. In this paper, we 
have used the conservative value, $\Gamma = 10^{-26} \mathrm{s^{-1}}$ 
for our research. There are many works where the constraints on the 
decay rate of gravitino are obtained. \cite{Huang:2011xr} used the Fermi observations 
of nearby galaxy clusters to get the constraints on this parameter. They found 
that the limits on the lifetimes of gravitino from the clusters 
observations is $\tau(1/\Gamma) \sim 2 \times 10^{26}s$ and it is mostly 
independent of on the dark matter mass. The limits obtained from the lines signals 
are different and the lifetime becomes smaller with the increasing of dark matter mass. 

For the future neutrino observations, such as KM3Net, because its effective area 
and volume will be improved, it is expected that the constraints on the 
abundance of UCMHs will be much more stronger.
\label{sect:discussion}

\normalem
\begin{acknowledgements}
We thanks Xiao-Jun Bi and Xiao-Yuan Huang for very usefull comments and suggestions. 
Yu-Peng Yang is supported by the 
National Science Foundation for Post-doctoral Scientists of China (Grant No.2013M540434) 
and the Special Funds of the National Science Foundation of China (Grant No.11347148). 
Ming-Zhe Li is supported in part by National Science
Foundation of China under Grants Nos. 11075074 and 11065004.

\end{acknowledgements}

\bibliographystyle{raa}
\bibliography{bibtex}

\end{document}